\documentclass{article}
\usepackage{amsmath,graphicx,mlspconf}
\usepackage{enumitem}
\usepackage{amssymb}
\usepackage{booktabs}
\usepackage{siunitx}

\copyrightnotice{U.S.\ Government work not protected by U.S.\ copyright}

\copyrightnotice{978-1-7281-6338-3/21/\$31.00 {\copyright}2021 Crown}

\copyrightnotice{978-1-7281-6338-3/21/\$31.00
{\copyright}2021 European Union}

\copyrightnotice{978-1-7281-6338-3/21/\$31.00 {\copyright}2021 IEEE}

\toappear{2021 IEEE International Workshop on Machine Learning for Signal Processing, Oct.\ 25--28, 2021, Gold Coast, Australia}

\title{CAESynth: Real-Time Timbre Interpolation and Pitch Control with Conditional Autoencoders}
\name{Aaron Valero Puche, Sukhan Lee}
\address{
        Artificial Intelligence School \\
        Sungkyunkwan University \\ 
        Suwon-si 16419, Republic of Korea
}

\begin{document}

\maketitle

\begin{abstract}
In this paper, we present a novel audio synthesizer, CAESynth, based on a conditional autoencoder. CAESynth synthesizes timbre in real-time by interpolating the reference sounds in their shared latent feature space, while controlling a pitch independently. We show that training a conditional autoencoder based on accuracy in timbre classification together with adversarial regularization of pitch content allows timbre distribution in latent space to be more effective and stable for timbre interpolation and pitch conditioning. The proposed method is applicable not only to creation of musical cues but also to exploration of audio affordance in mixed reality based on novel timbre mixtures with environmental sounds. We demonstrate by experiments that CAESynth achieves smooth and high-fidelity audio synthesis in real-time through timbre interpolation and independent yet accurate pitch control for musical cues as well as for audio affordance with environmental sound. A Python implementation along with some generated samples are shared online.
\end{abstract}
\begin{keywords}
Timbre Interpolation, Autoencoders, Disentanglement, Audio Synthesis, Audio Mixed Reality.
\end{keywords}
\section{Introduction}
\label{sec:intro}

A controllable synthesizer is a type of system capable of generating audio with a certain degree of control over the audio properties, namely timbre, pitch, velocity and so on. By controlling these properties, new cues can be elaborated or manipulated. More concretely, we focus on two main operations; pitch control and audio mixing. The first allows tone assignment whereas the latter permits the exploration of new timbres from existing ones. Both operations are non-mutually exclusive, i.e. both operations can be demanded simultaneously without interference. For this purpose, an ideal system should accomplish the following: (1) the independent factors presented in the data representation should be disentangled allowing custom control, (2) the output feature must be conditioned upon the input samples granting conditional mixture. In this work, we do not limit our scope to operate only with musical signals but also we intend to discover new sounds from environmental sounds. These cues can be helpful for mixed reality applications where the corresponding sounds emitted by daily life objects can be blend in order to provide more descriptive melodies \cite{Abboud2014EyeMusicIA}; for instance, an effective way of perceiving a dog could be creating a new instrument that sounds like a bark as referred to audio affordance.

\begin{figure}[t]
 \centering
 \includegraphics[width=8.5cm]{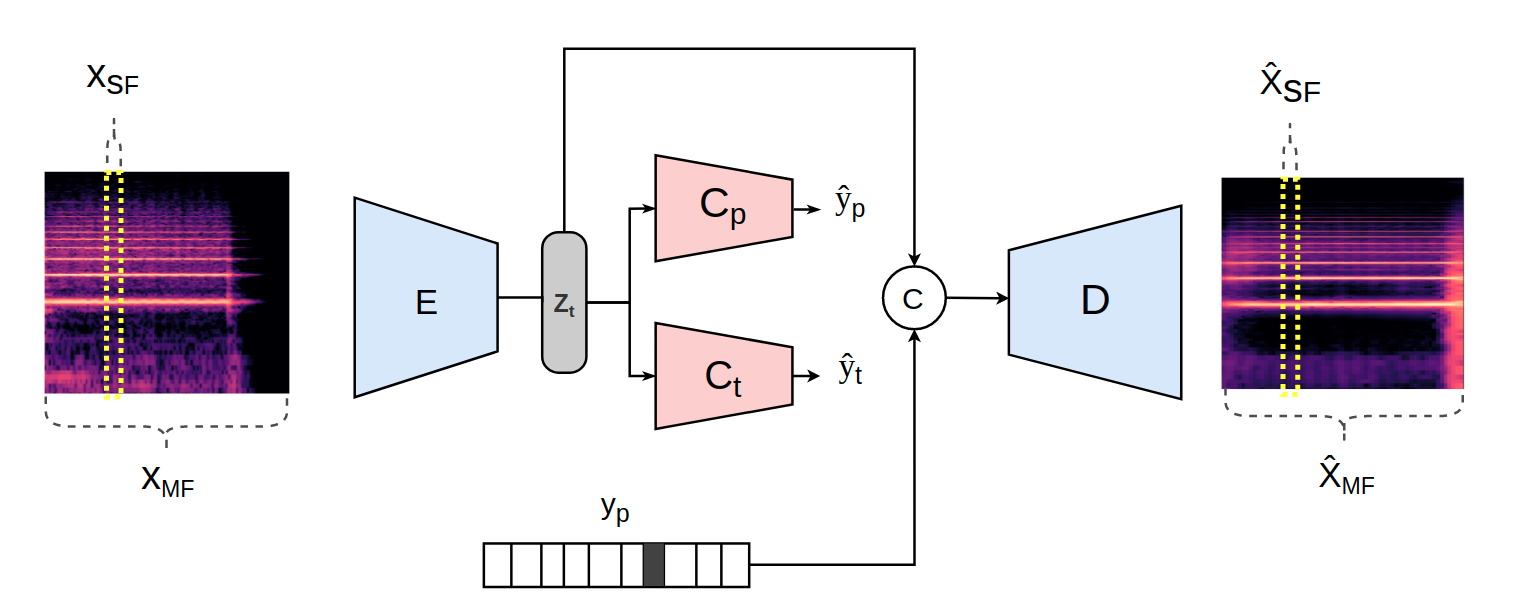}
  \centering
  \caption{\small Model architecture of CAESynth. The model is able to operate in single-frame ($X_{SF}$) and multi-frame ($X_{MF}$) modes. $E$ infers the timbre latent code $z_t$ which is then concatenated to the desired pitch $y_p$ and reconstructed by $D$. Two regularization terms are added in the latent space via neural classification. On one the hand $C_t$ ensures a correct clustering w.r.t timbre classes, whereas $C_p$ penalizes accurate pitch prediction via adversarial training.
  }
  \label{fig:caesynthdiag}
\end{figure}




The generation of audio is a challenging task due to the complex nature of the audio representation. This is mainly due to the great amount of samples per second, periodicity and time dependency. To deal with these challenges, the well-known Short-Time Fourier Transform (STFT) has been widely applied to elaborate a more compact representation by extracting the magnitude of frequency components over time, known as spectrogram (phase is typically discarded). The mentioned representation gains importance in the musical domain because of the deep relationship between musical qualities and the frequency components, where the pitch quality is often referred to the fundamental frequency ($F_0$) and the timbre quality can be determined by the harmonic content of the cue. On the other hand, environmental sound is generally considered as "unpitched" and in some cases it can be categorized as noise. The different nature between musical and environment sound signals presents a challenge that we address in the underneath sections in this paper.


\copyrightnotice{978-1-7281-6338-3/21/\$31.00 {\copyright}2021 IEEE}

\section{Related Work}


\textbf{Style Transfer}. Original works in style transfer \cite{gatys2016image} were introduced in computer vision. An iterative algorithm blends two input images, a content and a style images, the first provides the content or structural information whereas the second provides the overall style. An element-wise distance metric such as the mean square error (MSE) is used to define the content error from the iterated feature map and the content image. In contrast, the style error is defined as the Gramian matrix between the feature map and the style image. This technique has also been applied in the audio domain in \cite{grinstein2018audio}, although instead of using images, this method operates with the spectral information. Audio style transfer differs from our work in that style transfer mixture does not impose any disentanglement of the pitch quality.

\textbf{Domain Translation}. This problem aims at finding the mapping from a source domain to a target domain, each domain having different data distribution. Generative Adversarial Network (GAN) \cite{goodfellow2014generative} is a popular approach for this problem. GAN is an unsupervised neural algorithm for high-fidelity synthesis composed of two architectures competing with each other in a zero-sum game, the generator and the discriminator. 
A widely extended image-to-image translation method called CycleGAN \cite{CycleGAN2017}, defined the cycle-consistency loss objective to preserve the structure content of the input distribution yet converting relevant features to a second domain. Another similar work \cite{amodio2019travelgan} can carry out the image translation using the contrastive learning framework. The mentioned techniques have also been extended to the audio domain with little variations. A third generation CycleGAN especially designed for voice conversion is presented in \cite{kaneko2020cyclegan}. More related works addressing speech and musical tasks can be found in \cite{kameoka2019crossmodal,huang2019timbretron}. Despite domain translation methods have shown promising results at transferring certain data qualities, it is not commonly utilized for disentanglement or mixture tasks. 

\textbf{Timbre Interpolation.} Deep learning strategies were presented to generate new timbres via interpolation. The interpolation typically occurs in the latent space where the independent factors of the data distribution are more susceptible to be shaped. To this end, the timbre is often disentangled from other audio qualities, primarily the pitch. \cite{engel2017neural} proposes NSynth, a WaveNet-based \cite{oord2016wavenet} autoencoder that interpolates in the feature space which is extracted from raw audio signals. Also, they show evidence of disentanglement between pitch and timbre due to the conditioning. In \cite{colonel2020conditioning}, a similar work to ours, achieves single-frame timbre interpolation with pitch conditioning in both input space and latent space. Gaussian Mixture model combined with Variational Autoencoder GMVAE \cite{luo2019learning} was proposed in order to obtain effective disentangled latent code for timbre and pitch that allows timbre transfer. However, this method has not shown evidence of pitch control and has only been trained with a small dataset. Finally GANSynth was proposed in \cite{engel2019gansynth}, in this work the generator learns to produce STFT magnitude as well as the instantaneous frequency feature. The model generates high-fidelity melodies with accurate pitch control, nonetheless this method cannot infer the generated timbre since the timbre latent vector is drawn with random sampling. 

\section{Problem Definition and Approach}

In this paper, we intend to solve the following problems associated with sound synthesis as our main contributions: an effective way of generating a novel timbre with well-formed timbre clusters in the latent feature space based on interpolation of reference sounds, while controlling pitch of the synthesized timbre independently of reference sound pitch contents, where reference sounds include musical instruments as well as environmental sounds for audio affordance.

We define our model as a conditional autoencoder (CAE) with regularization and conditioning in the latent space. An autoencoder is a type of deep neural network composed of an encoder and a decoder, these two components are in charge of mapping the data from the input space to a typically compressed hidden space known as latent space. The regularization is characterized by two terms, an adversarial term to address pitch disentanglement and a classification term aiming to uniformly distribute the timbre quality. We find that using these regularization objectives is a key strategy to achieve reliable conditioning. The conditioning is carried out by concatenating the desired one-hot pitch embedding along with the inferred latent code. Finally, our method produces real-time synthesis, which is fundamental for certain applications. The regularization as well as the inference capability of the encoder makes autoencoders a good fit to fulfill the earlier mentioned requirements, (1) and (2) respectively.


More explicitly, \textbf{our contributions} are as follows:
\begin{itemize}[topsep=3pt, partopsep=0pt, noitemsep]
  \item Pitch control by selecting the desired pitch level in the latent space without altering the timbre of the input.
  \item Synthesis of new timbres via latent space interpolation.
  \item Exploring new sounds by combining environment and musical cues while keeping pitch control capabilities.
 \end{itemize}

\section{Datasets}

We consider two datasets, NSynth \cite{engel2017neural} and FreeSoundDataset (FSD) \cite{fonseca2020fsd50k} for training and testing our method. On the one hand, NSynth is a high quality musical instrument dataset constructed by the Magenta group at Google. Each sample is 4 seconds long sampled at 16 kHz and annotated with musical properties, such as pitch, instrument family, instrument identifier, velocity and other secondary features. The dataset contains 10 instrument family classes, i.e. \textit{bass}, \textit{brass}, \textit{vocal}, \textit{guitar}, \textit{organ}, \textit{keyboard}, \textit{reed}, \textit{string}, \textit{flute} and \textit{mallet}, as well as 3 different instrument source classes, namely \textit{acoustic}, \textit{synthetic} or \textit{electronic}. We define $N_t=28$ timbre classes resulting from the pair combination of available instrument family and source labels. Regarding the pitch classes, we select $N_p=61$ different pitch levels following MIDI standards. Furthermore, we create a new training and validation partition since the one originally provided does not consider instrument overlap between partitions. The dataset has a total of 207.030 samples for training and 22.988 for testing.

On the other hand FSD samples consist of variable length segments collected from multiple sources sampled at 44.1 kHz. The dataset contains a total of 51.197 segments and about 200 sound categories in a wide range of settings. Nevertheless, we trim the dataset by selecting a set of 38 unique classes, where each class represents a common entity in our daily lives e.g. door slamming, water dripping, air conditioning operating. A small set of samples per class is manually selected and cropped to a maximum duration of 4 seconds and resampled to 16 kHz for convenience.

\section{Method}

In this section, we describe the method as well as the details of the data representation and training. Fig~\ref{fig:caesynthdiag} shows the overall diagram of the proposed autoencoder model with latent regularization and pitch conditioning. 

\subsection{Data Representation}
\label{ssec:data_rep}

The input audio representation is obtained by first cropping the samples to 1 second long and normalizing in the range $[-1,1]$. Then the STFT magnitude is extracted from the time-based signal, followed by the absolute and log operators. STFT is computed with a window length of 2048 and a hop size of 256. Mel-filters have been extensively applied in the literature to emphasize the lower frequencies and to compress further the frequency axis resolution \cite{luo2019learning,prenger2018waveglow}, however, we observe that this operation rather harms the audio quality in our setup. The resulting input shape is $X_{MF} \in \mathbb{R}^{1024 \times 64}$ for multi-frame mode (MF), where 1024 corresponds to the frequency bins and 64 to the spectral frames, and $X_{SF} \in \mathbb{R}^{1024}$ as the single-frame (SF) representation.

The phase component is recovered back by the conventional approach Griffin-Lim algorithm \cite{Griffin1983SignalEF}. We choose 30 as the number of iterations in our experiments.

\subsection{Conditional Autoencoder Model}
\label{ssec:model}

Autoencoders are typically composed of two sequential stages, a feature extraction stage and a reconstruction stage. During the feature extraction, an encoder \textbf{$E$} projects the most relevant features of the input data to a high dimensional latent space. In order to assure a desirable clustering of timbre quality and disentanglement of pitch, two independent classifiers, \textbf{$C_t$} and \textbf{$C_p$} respectively are added. These classifiers take the inferred latent code \textbf{$z_t$} as input to predict the timbre \textbf{$\hat{y}_t$} and pitch \textbf{$\hat{y}_p$} classes. Finally, a one-hot embedding \textbf{$y_p$} corresponding to the desired pitch is concatenated to \textbf{$z_t$} and decoded with a symmetric decoder \textbf{$D$} to obtain the target STFT feature \textbf{$\hat{X}$}. Ideally, \textbf{$z_t$} should be easily classified in terms of timbre yet difficult to classify in terms of pitch, if these two conditions are satisfied, we can assume that a correct disentanglement w.r.t the pitch is achieved and decoder \textbf{$D$} totally relies on the pitch conditioning to reconstruct the desired pitch.

First of all, the neural model depicted in Fig~\ref{fig:caesynthdiag} can operate with either MF and SF input modes. Convolutional neural networks are chosen as the backbone of the MF architecture, due to its similarity with image representations. Inspired by \cite{engel2017neural}, we design the network $E$ with a stack of 10 convolutional layers of kernel size 4 and a stride of 2 until reducing the frequency-time axis completely. The channel axis is progressively expanded until reaching 512 channels. Directly after every layer, a LeakyReLU activation function and a batch normalization layer are added. $D$ follows a symmetric design w.r.t $E$ yet replacing the conv. layers by transposed conv. layers. As for the SF mode, we opt for fully-connected (FC) architecture, both $E$ and $D$ consist of 6 FC layers stacked with Tanh activations. The size of the latent code is 32. Finally, both pitch and timbre classifiers are built of 4 stacked FC layers with intercalated ReLU activations.

\subsection{Training}
\label{ssec:training}

Depending on the application, the model can be trained in supervised fashion with a musical dataset, where both the ground truth timbre and pitch labels are provided, or in semi-supervised fashion with musical and environmental sound datasets together. In the latter case, the environmental sound classes are considered as timbre classes and the pitch label remains unknown. We train the single and the multi frame models for 10 and 16 hours respectively, on a single Titan V GPU, batch size of 8 and learning rate $10^{-4}$.

When training the supervised models, we consider three loss objectives; reconstruction, timbre classification and pitch adversarial terms. The reconstruction term is computed as the weighted MSE between the input and the output and it is minimized by \textbf{$E$} and \textbf{$D$} networks. The weights follow a linearly decaying trend w.r.t the frequency axis, starting from a weight value of 10 at the lowest frequency bin until 1 at the last frequency bin. In addition, to make sure the input and the output have the same pitch quality, the ground truth pitch is used as pitch conditioning \textbf{$y_p$} during training. We define the second term as the timbre classification loss that is minimized by both \textbf{$E$} and \textbf{$C_t$}. Finally, the last objective is defined by the pitch classification which is minimized by \textbf{$C_t$} and maximized by \textbf{$E$}. Similarly to a discriminator in GAN framework, while \textbf{$C_p$} aims to find the pitch information in \textbf{$z_t$}, \textbf{$E$} aims to discard the pitch information in a zero-sum game. Both timbre classification and pitch adversarial losses are defined as multi-class cross-entropy loss. The objectives are as follows:
\begin{align*}
\mathcal{L}_{E,D,C_t} = {}& \mathbf{E}[W(X-\hat{X})^2] - \mathbf{E}[\sum\limits_{i=1}^{N_t}y_{t,i} \log(\sigma(\hat{y}_{t,i}))] \\ {}& \qquad \quad \qquad \qquad + \mathbf{E}[\sum\limits_{i=1}^{N_p} y_{p,i}\log(\sigma(\hat{y}_{p,i}))]
\\
\mathcal{L}_{C_p} = {}& -\mathbf{E}[\sum\limits_{i=1}^{N_p} y_{p,i}                      \log(\sigma(\hat{y}_{p,i}))]
\end{align*}
where $W$ is the linearly decaying weight array and $\sigma()$ is the softmax function.

In the case of semi-supervised training, the objectives remain the same only when forwarding instrumental samples, yet removing the pitch adversarial term when an environmental sample is given as the pitch label is unknown. Besides, we augment the FSD samples using random pitch shift with 50\% probability no more than one octave upwards and downwards. Spectral visualizations and pitch shifting are carried out with the open-sources Python library Librosa \cite{mcfee2015librosa}.

\begin{figure}[t]
 \centering
 \hspace*{-0.5cm}\includegraphics[width=8.5cm]{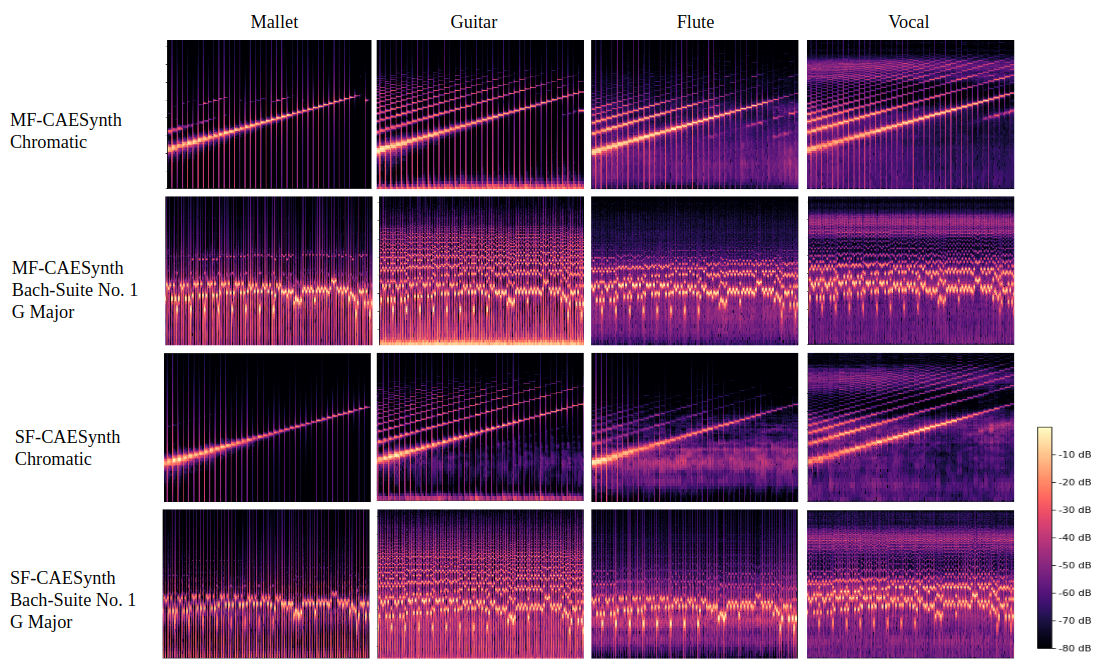}
  \centering
  \caption{ \small Spectral synthesis plots of chromatic scale and Bach Suite No 1. in G major generated from a single input sample. We compare the MF and SF outputs for four different instrument families.
  } 
  \label{fig:pitchcontrol}
\end{figure}

\section{Evaluation}

We compare both implementations, single-frame and multi-frame, with two baselines. The first baseline was originally introduced in \cite{engel2017neural}, it is a spectral multi-frame AE with pitch conditioning like our model, however, this method differs from our method in that it does not include regularization. The second baseline is presented in \cite{colonel2020conditioning}. It consists of a single-frame model with pitch conditioning at the input and latent space. The quantitative metrics are as follows:

\begin{itemize}[topsep=3pt, partopsep=0pt,noitemsep]
  \item \textbf{MSE}: error between normalized inputs and outputs when using the ground truth pitch as conditioning.
  \item \textbf{Speed}: the time spent to synthesize 1 second of audio.
  \item \textbf{Latent Pitch Accuracy (LPA)}: a fully-connected based classifier (identical to $C_p$ and $C_t$) is trained to predict pitch class given $z_t$ from pretrained models. A lower classification loss can show disentanglement evidence between the pitch quality from other qualities.
  \item \textbf{Synthesis Pitch Accuracy (SPA)}: similarly to the Inception Score \cite{salimans2016improved,engel2019gansynth}, we train a multi-frame classifier (identical to $E$) with NSynth dataset to evaluate synthetic samples (model's output), yet replacing the pitch conditioning with a randomly sampled pitch embedding. Intuitively, a high accuracy result means an accurate capability to synthesize the desired pitch. The pretrained classifier obtains 94.84\% and 92.12\% pitch accuracy in the training and testing sets respectively.
 \end{itemize}
 
In sections~\ref{ssec:recons},~\ref{ssec:pitch_control} and~\ref{ssec:tim_interp}, the methods are evaluated in a supervised manner only with NSynth musical dataset. Afterwards in section \ref{ssec:env_mix}, we discuss the semi-supervised model trained with NSynth and FSD datasets.

\subsection{Reconstruction}
\label{ssec:recons}

The reconstruction is evaluated in terms of synthesis speed and MSE distance between inputs and outputs. In regard to the generation speed, the results show that SF generates approximately 1200 times faster than real-time in contrast with 370 in the case of MF models. As a result, we can conclude that these methods can effectively be utilized for real-time applications. Identity reconstruction is found to be lower when using the baselines models, this can be due to the imposed latent regularization during training. Moreover, SF has empirically lower reconstruction error than MF mode.

\subsection{Pitch control}
\label{ssec:pitch_control}

To achieve controllable pitch synthesis, two conditions must be satisfied; first, the encoder must discard all correlation with pitch quality, so that other qualities can be more accordingly encoded. Secondly, the decoder must rely only on the pitch conditioning to synthesize the desired pitch. These two conditions are translated into low LPA and high SPA. Table \ref{Tab:results} shows that the proposed model generates substantially more uncorrelated latent codes than the baselines for both modes, especially the single-frame model, which obtains the lowest accuracy (1.62\%). Not only that, but also the SPA metric confirms that the proposed models synthesize samples with pitch conditioning with high accuracy in contrast to the baselines which barely achieve 27.6\% in the single-frame case.

As a qualitative result, Fig. \ref{fig:pitchcontrol} depicts some pitch control examples synthesized from our models given a single instrumental sample. The outputs consist of chromatic scales and a classical composition, i.e. Suite N°1 Prelude in G in major by J.S.Bach. In the chromatic case, we simply start off by the minimum available pitch level in the dataset and successively increase the level by one step. As for the melody, we obtain the temporal and pitch information from open-source files in MIDI format. As shown in the figure, both models are able to generate the target melodies regardless of the instrument. Besides, it is important to notice that the input timbre remains unaltered while selecting different pitch embeddings. 

\begin{figure}[t]
 \centering
 \hspace*{-0.25cm}\includegraphics[width=8.5cm]{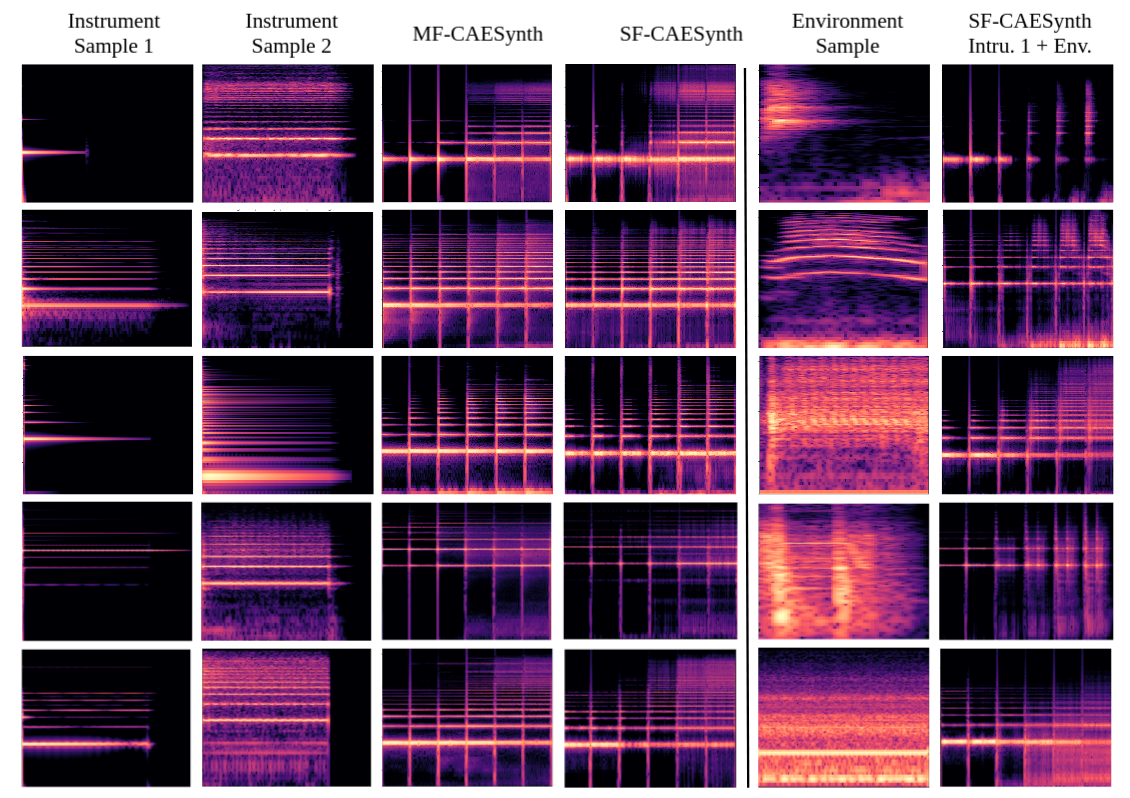}
  \centering
  \caption{\small Spectral plots of progressive timbre interpolation between
instrumental samples ($3^{rd}$ and $4^{th}$ column) and instrumental with environmental
samples ($6^{th}$ column). The environment samples are mixed with instrument sample 1.}
  \label{fig:timbreinterp}
  
\end{figure}

\subsection{Timbre Interpolation}
\label{ssec:tim_interp}

\begin{figure}[!ht]
 \centering
 \includegraphics[width=8.5cm]{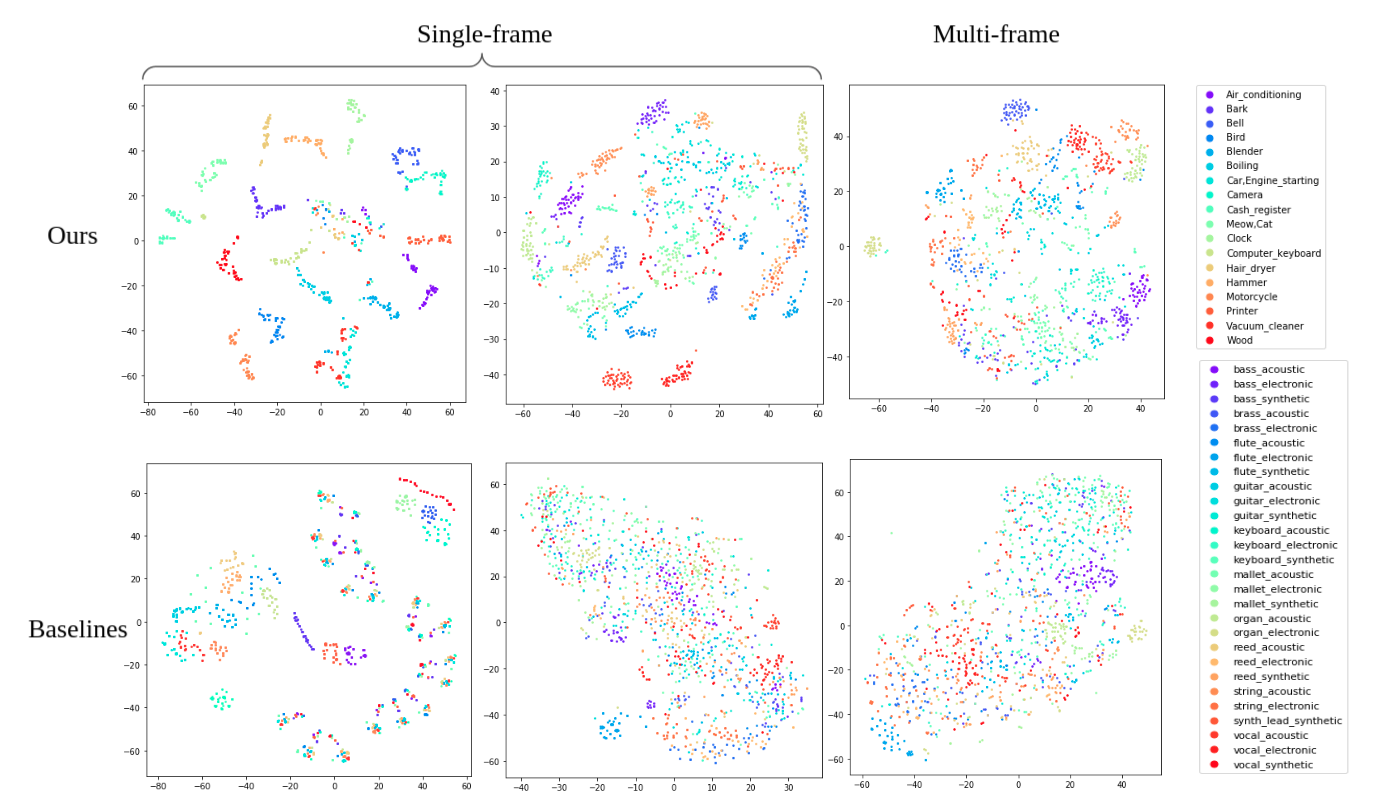}
  \centering
    \caption{\small t-sne plots of $z_t$ distributed by timbre classes of environment ($1^{st}$ column) and musical samples ($2^{nd}$ and $3^{rd}$ column)}
  \label{fig:ztdist}
\end{figure}

Clustering is frequently used to describe the relationship of different groups within the data. In our scope, a better timbre clustering can help us to better geometric arrangement of the timbre factors. In order to visualize the feature space distributed by instrument families, t-sne \cite{van2008visualizing} is utilized for reducing the high-dimensional latent space to a 2D space. Fig. \ref{fig:ztdist} shows how the proposed models are able to form clusters better than the baselines as a result of the regularization.

\setlength{\heavyrulewidth}{1.5pt}
\setlength{\abovetopsep}{4pt}
\setlength{\tabcolsep}{5pt}
\begin{table}
\centering
\caption{\small Model's quantitative comparison.}
\tiny
\begin{tabular}{*6c}
\toprule
Mode &  Method & MSE & Speed (ms) & LPA (\%) & SPA (\%)\\
\midrule
Single-Frame  & ours   & $7.36 \cdot 10^{-3}$ & \textbf{0.803} $\boldsymbol{\pm}$ \textbf{0.17} & \textbf{1.62} & \textbf{95.84} \\
{}&  baseline \cite{colonel2020conditioning} & $\mathbf{5.0 \cdot 10^{-3}}$  &{} & 93.16 & 27.60 \\
\midrule
Multi-Frame &  ours & $15.31 \cdot 10^{-3}$ & 2.70$\pm$ 0.58  & \textbf{25.03} & \textbf{89.46} \\
{}   &  baseline \cite{engel2017neural}  & $\mathbf{9.8 \cdot 10^{-3}}$  & {}  & 92.04 & 0.33 \\
\bottomrule
\end{tabular}
\label{Tab:results}
\end{table}

A set of latent codes can simply be interpolated as follows: $z_{intp} = \sum_{i=0}^K w_i z_{t,i}$, where K is the number of embeddings to interpolate and $w_i$ is the weight contribution of $ith$ sample. This interpolation can be accomplished even when the input samples have different pitch values as the pitch factors should not be present in $z_t$. Fig. \ref{fig:timbreinterp} exhibits some timbre interpolation examples between different instruments and pitch levels. The interpolation is performed progressively in time, or more formally, the instrument weights take values $w_1 = \alpha $ and $w_2= 1- \alpha$ where $\alpha$ is shifted from 0 to 1 over time. We observe that there is a smooth transition between both instruments while the pitch is constant.

\subsection{Mixing Musical And Environmental Sounds}
\label{ssec:env_mix}

FSD and NSynth datasets are significantly different in nature. As depicted in Fig.\ref{fig:timbreinterp}, the musical samples tend to be constant in time (x-axis), whereas the vast majority of environmental sounds are highly dependent on time e.g cash cashier, car engine etc. While training the MF model in a semi-supervised manner, we noticed that the model fails to reconstruct both musical and environmental sounds precisely, but rather tries to generate STFT features similar to the musical data, i.e. constant in time. This issue is mitigated when utilizing the SF model as each frame is mixed independently, thus the difference in nature does not affect the interpolation. Although the model is trained by default with constant pitch conditioning $y_p=0$ given FSD samples, the reconstruction is still recognizable when modifying the conditioning to a different embedding. Moreover, we notice that the reconstructed sample highlights the pitch component and that the frequencies of the cues are shifted according to the selected conditioning. Fig.\ref{fig:timbreinterp} illustrates some musical and environmental sound mixture examples. Audio based samples are publicly shared online.

\section{Conclusions}
In this paper, we presented CAESynth, a model for high-fidelity controllable synthesis based on conditional autoencoder in real-time. We demonstrated that the presented work can successfully synthesize complex melodies given a single sample, since the pitch can be controlled via latent space conditioning while remaining the instrument's identity (timbre) unchanged. Not only that, but also this method allows us the synthesis of new interpolated timbres on demand. Furthermore, We discussed that the latent space regularization with classification and adversarial classification are key strategies to ensure a proper distribution in the feature space. Lastly, but not least importantly, we extended our method to perform audio interpolation between musical and non-musical cues. By smoothly mixing these two domains, we believe that the model can generate melodic sound with description of daily environments, aiming to innovate in the mixed reality field.

\section{Acknowledgement}
This research was funded, in part, by "Deep Learning Based Cross-Sensory Transfer for Visually Impaired” Project of National Research Foundation (NRF) of Korea $\\$ (2020R1A2C200956811), and, in part, by AI Graduate School Program, Grant No. 2019-0-00421, and by ICT Consilience Program, IITP-2020-0-01821, of the Institute of Information and Communication Technology Planning \& Evaluation (IITP), sponsored by the Korean Ministry of Science and Information Technology (MSIT).




\bibliographystyle{IEEEbib}
\bibliography{References}

\end{document}